\documentclass[aps,twocolumn,pra,footinbib,floatfix,showpacs,showkeys,
               superscriptaddress,eqsecnum]{revtex4}
\usepackage{epsf,color,colordvi} 
\usepackage{graphicx}
\usepackage{amsmath}
\newcommand{\lapproxeq}{\mathrm{\raisebox{-0.6ex}{$\stackrel{\textstyle <}{\sim}$}}}

\begin{document}

\title{Multi-phase matching in the Grover algorithm}

\author{F.M. Toyama}
\email{toyama@cc.kyoto-su.ac.jp}
\affiliation{Department of Information and Communication Sciences,
  Kyoto Sangyo University, Kyoto 603-8055, Japan}
\author{W. van Dijk}
\email{vandijk@physics.mcmaster.ca}
\affiliation{Physics Department, Redeemer University College,
  Ancaster, Ontario L9K 1J4, Canada}
\affiliation{Department of Physics and Astronomy, McMaster University,
  Hamilton, Ontario L8S 4M1, Canada}
\author{Y. Nogami}
\email{nogami@mcmaster.ca}
\affiliation{Department of Physics and Astronomy, McMaster University,
  Hamilton, Ontario L8S 4M1, Canada}
\author{M. Tabuchi}
\affiliation{Department of Information and Communication Sciences,
  Kyoto Sangyo University, Kyoto 603-8055, Japan}
\author{Y. Kimura}
\affiliation{Department of Information and Communication Sciences,
  Kyoto Sangyo University, Kyoto 603-8055, Japan}

\date{\today}

\begin{abstract}

  Phase matching has been studied for the Grover algorithm as a way of
  enhancing the efficiency of the quantum search.  Recently Li and Li
  found that a particular form of phase matching yields, with a single
  Grover operation, a success probability greater than 25/27 for
  finding the equal-amplitude superposition of marked states when the
  fraction of the marked states stored in a database state is greater
  than 1/3. Although this single operation eliminates the oscillations
  of the success probability that occur with multiple Grover
  operations, the latter oscillations reappear with multiple
  iterations of Li and Li's phase matching.  In this paper we
  introduce a multi-phase matching subject to a certain matching rule
  by which we can obtain a multiple Grover operation that with only a
  few iterations yields a success probability that is almost constant
  and unity over a wide range of the fraction of marked items.  As an
  example we show that a multi-phase operation with six iterations
  yields a success probability between 99.8\% and 100\% for a fraction
  of marked states of 1/10 or larger.

\end{abstract}



\maketitle

\section{Introduction}
\label{sec:intro}
The quantum search algorithm introduced by
Grover~\cite{grover96,grover97a,grover97b,grover01} constitutes a
major advance in quantum computing. It enables us to find a marked
state stored in a database state consisting of $N$ unordered basis
states in only ${\cal O}(\sqrt{N})$ Grover operations.  A number of
modifications and generalizations of the original Grover search
algorithm have been
proposed~\cite{grover98,long99,long01,collins02,tulsi06,li07,younes07}.
In particular, phase matching methods in the Grover algorithm have
been extensively examined~\cite{long99,long01,li07}.  The outcome of
the search algorithm is characterized in terms of $P(\lambda)$, the
probability of obtaining an equal-amplitude superposition of the
marked states where $\lambda$ is the ratio of the marked states to all
the states stored in the original database state.

Recently, Li and Li~\cite{li07} proposed a new phase matching for the
Grover algorithm and they obtained an improved success probability
$P(\lambda)$ over a wide range of the ratio $\lambda$.  They
introduced the set of the Grover operators (details are described in
Eqs.~(\ref{eq:1}) and (\ref{eq:2})): $U =
I-(1-e^{i\alpha})\sum_{l=0}^{M-1}|t_l\rangle\langle t_l|$ and $V =
Ie^{i\beta}+(1-e^{i\beta})|0^{\otimes n}\rangle \langle 0^{\otimes
  n}|$.  The phase factor $e^{i\beta}$ in the first term of the
operator $V$ was first introduced in Ref.~\cite{li07}.  In the new
phase matching the number of phases is the same as the usual one but
the form of the phase shift operator $V$ is different.  Li and Li
found the remarkable result that \emph{a single Grover operation} of
the new phase matching yields $P(\lambda)>25/27$ for
$1/3\leq\lambda\leq 1$. This is significant in the sense that with
\emph{only one Grover operation} the efficiency of the Grover
algorithm is substantially improved in the range of values of
$\lambda$ where the efficiency of the original algorithm deteriorates.

This phase matching has another interesting aspect that was not
explicitly pointed out by Li and Li~\cite{li07}. For a given values of
$\lambda$ in the range $1/4\leq\lambda \leq 1$, one Grover operation
with the phases $\alpha=-\beta=\arccos(1-1/2\lambda)$ yields exactly
$P=1$. [See Eq.~(\ref{eq:12a}) in the following.]  This results was
obtained earlier by Chi and Kim~\cite{chi97} who considered a modified
Grover operator of arbitrary phase.  The special case of $\lambda=1/2$
yields $\alpha=-\beta=\pm\pi/2$, which are the phases found in
Ref.~\cite{li07}.  This aspect of the phase matching is also
significant because it implies that one can always find the
equal-amplitude superposition of the marked states by only one Grover
operation when $\lambda$ is greater than 1/4 by tuning the phases
$\alpha$ and $\beta$ appropriately for the given $\lambda$.
Conditions for a success probability of unity have been studied by
previous authors.  See, for example, Refs.~\cite{hoyer00,long01a}.

It should be pointed out, however, that the so-called new phase
matching of Ref.~\cite{li07} is equivalent to the original phase
matching of Long \emph{et al.}~\cite{long99}.  When the second
operator is defined as $V'=e^{-i\beta}V$, it becomes the
phase-matching operator of Long \emph{et al.}  The only difference
between the two is that the overall state is multiplied by a phase
factor and so the amplitudes of the components are different, but the
probabilities are the same.  Thus the remarkable result of Li and Li
can also be seen to follow from the operator of Long \emph{et al.}
Analytically the formulation by Li and Li is somewhat more transparent
and hence we use it throughout this paper, except in the Appendix
where we explicitly show the equivalence of the two formulations by
calculating the probability profile.

Thus a number of aspects of the Grover algorithm with phase matching,
already alluded to, are of particular interest and they form the
objectives of this study.  We focus on high success probabilities with
as few iterations as possible in order to enhance the efficiency of
the quantum search.  We emphasize the following three objectives: (1)
the elucidation of features of the phase-matched Grover operations
with a small number of iterations that yield success probabilities
$P(\lambda)$ close to one over a wide range of values of $\lambda$,
(2) given a value of $\lambda$ the determination of the phase-matched
Grover operator(s) that results in $P(\lambda)=1$ exactly, and (3) the
elucidation of the features of the phase-matched Grover operators that
allow us to obtain $P(\lambda)=1$ for very small values of $\lambda$.

In this paper we explore the search algorithm with these objectives in
mind using the advantages of a few multiple Grover operations with
phase matching.  It is well known that a multiple application of the
original Grover operation gives rise to intensive oscillations of $P$
as a function of $\lambda$ and such oscillations deteriorate the
efficiency of the algorithm.  This undesirable feature remains even in
the new phase matching of Li and Li, as we will illustrate.  We show
that if we introduce a \emph{multi-phase} matching subject to a
certain matching rule, we can obtain a multiple Grover operation that
yields a success probability almost constant and unity over a wide
range of $\lambda$, e.g., $0.1 \leq \lambda\leq 1$.  This is also
significant in the sense that when $\lambda$ is greater than a small
minimum value we can always find the superposition of the marked
states with high degree of certainty without (re)tuning the phases.

In the next section we set up the algorithm of the multi-phase
matching in the framework of the phase matching of Li and
Li~\cite{li07} and analyze the efficiency of the algorithm by
considering a single matched phase and a two-stage multi-phase
matching.  We also obtain an exemplar of a good probability profile
for a six-stage multi-phase matched operator.  In
Sec.~\ref{k_iterations} we consider the success probability for small
$\lambda$ by using the Grover operations with a phase other than
$\pi$. We summarize our results in Sec.~\ref{summary}.

\section{Multi-phase matching in the framework of the new phase
  matching}
\label{multiphase}
The new phase matching in the Grover algorithm proposed by Li and
Li~\cite{li07} is defined with the two operators,
\begin{equation}\label{eq:1}
U=I-(1-e^{i\alpha})\sum_{l=0}^{M-1}|t_l\rangle\langle t_l|  
\end{equation} 
\begin{equation}\label{eq:2}
  V=Ie^{i\beta}+(1-e^{i\beta})|0^{\otimes n}\rangle\langle 0^{\otimes n}|.  
\end{equation} 
where $|0^{\otimes n}\rangle$ is the $n$-qubits initial state, $M$ is
the number of target (marked) states stored in an unstructured
database state, and the $|t_l\rangle$ denote the target or
marked states.  The database state is given as $|\phi\rangle =
H^{\otimes n}|0^{\otimes n}\rangle$, where $H$ is the Walsh-Hadamard
transformation. The state $|\phi\rangle$ is an equally-weighted
superposition of the $N=2^n$ basis states, $|w_l\rangle, \ l =
0,\dots,N-1$. The fraction $\lambda$ of the target states is defined
as $\lambda = M/N$. The $U$ and $V$ of Eqs.~(\ref{eq:1}) and
(\ref{eq:2}) are both unitary as was shown in Ref.~\cite{li07}. With
$\alpha=\beta=\pi$, $U$ and $V$ reduce to the Grover operators of the
original algorithm. As we mentioned in Sec.~\ref{sec:intro}, Li and Li
showed explicitly that \emph{a single Grover operation} of the new
phase matching $(H^{\otimes n}VH^{\otimes n})UH^{\otimes n}|0^{\otimes
  n}\rangle$ with $\alpha = -\beta=\pi/2$ yields a success probability
$P(\lambda)>25/27$ for $1/3\leq\lambda\leq 1$.

We introduce a multi-phase matching within the framework of the new
phase matching. We rewrite the database state $|\phi\rangle =
H^{\otimes n}|0^{\otimes n}\rangle = N^{-1/2}\sum_{l=0}^{N-1}
|\omega_l\rangle$ in terms of $\lambda$ as
\begin{eqnarray}\label{eq:3}
  |\phi\rangle = \frac{1}{\sqrt{N}}\sum_{l=0}^{N-1}|\omega_l\rangle 
               & = & \sqrt{\frac{N-M}{N}}|R\rangle + 
               \sqrt{\frac{M}{N}}|T\rangle \nonumber \\
               & = & \sqrt{1-\lambda}|R\rangle + \sqrt{\lambda} |T\rangle,
\end{eqnarray}
where
\begin{equation}\label{eq:4}
  |R\rangle = \frac{1}{\sqrt{N-M}}\sum_{l=0}^{N-M-1}|r_l\rangle,
  \ \ |T\rangle = \frac{1}{\sqrt M}\sum_{l=0}^{M-1}|t_l\rangle.
\end{equation}
The state $|T\rangle$ is the uniform superposition of the marked
states and $|R\rangle$ is that of the remaining states $|r_l\rangle$.
They are both normalized to unity and orthogonal to each other. In the
following, for convenience, we work in the two-dimensional space
defined by the basis $\{|R\rangle, |T\rangle\}$. The two-dimensional
representations of $U$ and $H^{\otimes n}VH^{\otimes n} = Ie^{i\beta}
+ (1-e^{i\beta})|\phi\rangle\langle\phi|$ are
\begin{widetext}
\begin{equation}\label{eq:5}
  U:\begin{pmatrix} 1 & 0 \\ 0 & e^{i\alpha} \end{pmatrix},
 \ \ \
  H^{\otimes n}VH^{\otimes n} : 
    \begin{pmatrix}
      (1-e^{i\beta})(1-\lambda)+e^{i\beta} & (1-e^{i\beta})
      \sqrt{\lambda(1-\lambda)} \\
      (1-e^{i\beta})\sqrt{\lambda(1-\lambda)} & (1-e^{i\beta})\lambda 
      + e^{i\beta} \end{pmatrix}.
\end{equation}
We write the multiple Grover operation with the multiple phases
$\alpha_j$ and $\beta_j$ $(j=1,\dots,k)$ as
\begin{equation}\label{eq:7}
  \begin{pmatrix} u_k \\ d_k \end{pmatrix} 
  = G(\alpha_k,\beta_k)G(\alpha_{k-1},\beta_{k-1})\cdots G(\alpha_1,\beta_1)
\begin{pmatrix} \sqrt{1-\lambda} \\ \sqrt{\lambda} \end{pmatrix},
\end{equation}
where one Grover operation
$G(\alpha_j,\beta_j)$ $(j=1,\dots,k)$ in this representation is  
\begin{equation}\label{eq:8}
  G(\alpha_j,\beta_j)= 
\begin{pmatrix}
      (1-e^{i\beta_j})(1-\lambda) + e^{i\beta_j} &
      (e^{i\alpha_j} - e^{i(\alpha_j+\beta_j)})\sqrt{\lambda(1-\lambda)} \\
      (1-e^{i\beta_j})\sqrt{\lambda(1-\lambda)} &
      (e^{i\alpha_j} - e^{i(\alpha_j+\beta_j)})\lambda + e^{i(\alpha_j+\beta_j)}
      \end{pmatrix}. 
\end{equation}
\end{widetext}
The success probability of finding the superposition of target states
is given by $P_k(\lambda)\equiv|d_k|^2$.

We now consider the one- and two-pair-phase cases before increasing
the phase-matching to six different pairs of phases in order to obtain
$P(\lambda)$ nearly equal to unity over a large range of values of
$\lambda$.  In other words, we discuss the $k=1$ and the $k=2$ cases
in detail first, and then proceed to the numerical results of the
$k=6$ case.

\subsection{Multi-phase matching with one pair of phases}
\label{one}
When $k=1$, Eq.~(\ref{eq:7}) reduces to
\begin{equation}\label{eq:10}
  \begin{pmatrix}u_1 \\ d_1 \end{pmatrix} =  G_1(\alpha,\beta)
  \begin{pmatrix}\sqrt{1-\lambda} \\ \sqrt{\lambda} \end{pmatrix}.
\end{equation}
Since we first focus on cases of complete success $P=|d_1|^2 = 1 -
|u_1|^2 = 1$, we can equivalently consider the condition $u_1 = 0$.
In general
\begin{equation}\label{eq:11}
  u_1 = \sqrt{1-\lambda} \, [1-\lambda+e^{i\beta}\lambda + (e^{i\alpha}
    -e^{i(\alpha+\beta)})\lambda \, ].
\end{equation}
The condition that $u_1$ be zero leads to
\begin{eqnarray}\label{eq:12}
  \frac{1}{\lambda} & = & 1 -\cos\alpha -\cos\beta 
  - \cos(\alpha+\beta) \nonumber \\ 
             && \ \ + i[\sin{(\alpha+\beta)}-\sin\alpha-\sin\beta].
\end{eqnarray}
The fact that $\lambda$ must be real implies that (1) $\beta =
-\alpha$, (2) either $\alpha$ or $\beta$ are zero, or (3) both
$\alpha$ and $\beta$ are zero. When $\beta =0$ in the operator $V$ of
Eq.~(\ref{eq:2}), the operator is the identity and the overall effect
of operator $U$ of Eq.~(\ref{eq:1}) by itself would cause the phase of
the marked states to be changed, but the probabilities of marked and
unmarked states would remain the same.  When $\alpha=0$, then $U=I$
and $G=H^{\otimes n} VH^{\otimes n}$.  The initial state
$|\phi\rangle$ is an eigenvector of $G$ with eigenvalue 1.  Thus $G$
does not cause any evolution in $|\phi\rangle$.  The success
probability is $P=\lambda$, which is the success probability of the
classical algorithm.  As no quantum improvement to the search
algorithm is achieved, we eliminate the case of $\alpha=0$ and any
nonzero $\beta$ from the solutions of Eq.~(\ref{eq:12}).  Thus only
the solution $\beta=-\alpha$ is meaningful, and yields, as mentioned
in Sec.~\ref{sec:intro} and in Ref.~\cite{chi97}, $P=1$ when 
\begin{equation}\label{eq:12a}
\alpha = -\beta = \arccos(1-1/2\lambda).
\end{equation}
Since $\lambda$ lies between zero and one, the range of $\alpha$ is
$\pi/3 \leq \alpha \leq \pi$. The boundary point of this range
$\alpha=\pi/3$ occurs when $P(\lambda=1)=1$, and similarly
$\alpha=\pi$ when $P(\lambda=1/4)=1$.

\begin{figure}[ht]
  \centering
   \resizebox{3.5in}{!}{\includegraphics{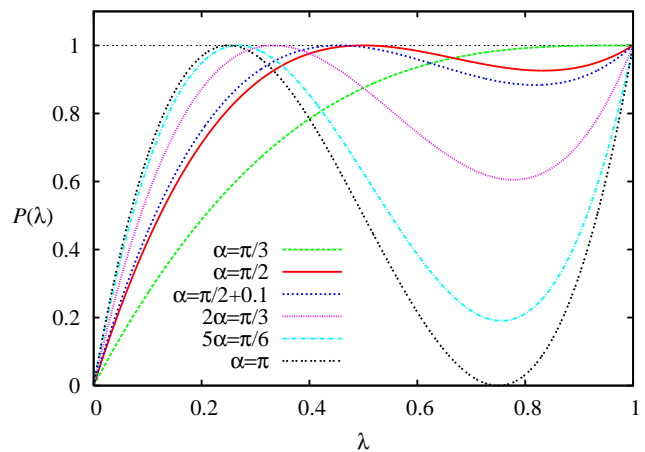}}
   \caption{(Color online) Plots of $P(\lambda)$ for different values of the
     parameter $\alpha$ for the case of one iteration. (See
     Eq.~(\ref{eq:13}).)}
\label{fig:1}
\end{figure}

We can express $P(\lambda)$ as a function of $\lambda$ depending on
the parameter $\alpha$,
\begin{eqnarray}\label{eq:13}
  P(\lambda) & = & 1 - |u_1|^2 \nonumber \\
             & = & \lambda \, [5-4 \, \cos\alpha 
      - 4 \, (1-\cos\alpha) \, (2-\cos\alpha) \, \lambda \nonumber \\
             && \ \ \ \             +4 \, (1-\cos\alpha)^2\lambda^2].
\end{eqnarray}
For $\alpha = \pi/2$ the equation reduces to Eq.~(14) of Li and
Li~\cite{li07} . Since Eq.~(\ref{eq:13}) is cubic in $\lambda$ we
expect a local maximum and a local minimum in the range $0<\lambda\leq
1$ at $\lambda_\mathrm{max}$ and $\lambda_\mathrm{min}$ respectively,
where
\begin{equation}\label{eq:14}
  \lambda_\mathrm{max}=\frac{1}{2 \, (1-\cos\alpha)}, \ \ 
  \lambda_\mathrm{min} = \frac{5-4\cos\alpha}{6 \, (1-\cos\alpha)}.
\end{equation} 
Furthermore the extrema are
\begin{equation}\label{eq:15}
  P(\lambda_\mathrm{max}) = 1, \ \ P(\lambda_\mathrm{min}) = 
  \frac{(1+\cos\alpha)(5-4\cos\alpha)^2}{27 \, (1-\cos\alpha)}.
\end{equation}
We illustrate different cases in Fig.~\ref{fig:1}.
It is evident that the $\alpha=\pi/2$ case, which is the one used by
Li and Li~\cite{li07}, gives the optimal profile for the success
probability.  Optimal here could be defined as the largest average $P$
over the range of $\lambda$, or the largest range of $\lambda$ over
which $P\geq 25/27$.

\subsection{Multi-phase matching with two pairs of phases}
\label{two}
We now consider Eq.~(\ref{eq:7}) for $k=2$ and we again concentrate on
the upper component of the vector $(u_2,d_2)^T$.  The general
expression for it is too lengthy to give here, but again we demand
that for an arbitrary value of $\lambda$ the imaginary part is zero to
obtain the matching relationship for the phases.  Apart from a factor
of $\sqrt{1-\lambda}$ the expression of $\mathrm{Im} \, u_2$
contains a term in $\lambda$ and another in $\lambda^2$.  Demanding
that the coefficients of each power of $\lambda$ vanishes gives us two
equations involving $\alpha_1$, $\alpha_2$, $\beta_1$, and $\beta_2$.
Solving for $\beta_1$ and $\beta_2$ in terms of $\alpha_1$ and
$\alpha_2$ we obtain the following four solutions:
\begin{equation}\label{eq:16}
\left.
\begin{array}{l}
  \{\beta_1=-\alpha_1, \ \beta_2 = 0\} \nonumber \\
  \{\beta_1=-\alpha_2, \ \beta_2 = -\alpha_1\} \nonumber \\
  \{\beta_1 = \beta_2 = 0\} \nonumber \\  
  \{\beta_1 = 0, \ \beta_2 = -(\alpha_1 + \alpha_2)\}.
\end{array}
\right.
\end{equation}
Since one of $\beta_1$ and $\beta_2$ is zero for the first and last
solution, the operation is then reduced to one iteration, and for the
third solution the two iterations would not change the probabilities
of the marked and unmarked states.  Thus the only solution that gives
new information is the one where $\beta_1=-\alpha_2$ and
$\beta_2=-\alpha_1$.  (The fact that $\mathrm{Im} \, u_2 = 0$ is a
necessary, but not a sufficient, condition for this solution.)  After
obtaining the matched phases for which $\mathrm{Im} \, u_2 =0$, we set
$\mathrm{Re} \, u_2 = 0$ to solve for the values of $\lambda$ which
gives $P=1$.

\begin{figure}[h]
  \centering
   \resizebox{3.5in}{!}{\includegraphics{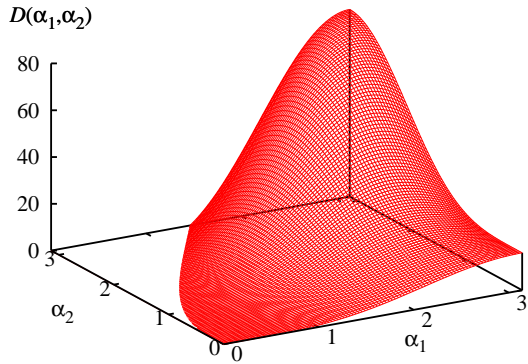}}
   \caption{(Color online) The discriminant of Eq.~(\ref{eq:17a}) as a
     function of $\alpha_1$, and $\alpha_2$.  The curve of
     intersection of the surface with the $D=0$ plane
     separates the smaller values of $\alpha_1$ for which there are no
     physical operators.}
\label{fig:3a}
\end{figure}

The expression for $u_2$ is then real and can be written as
\begin{eqnarray}\label{eq:17}
  u_2 & = & \{1+ 2\, [(1-\cos\alpha_1)(-2+\cos\alpha_2)
  -\sin\alpha_1\sin\alpha_2]\, \lambda \nonumber \\
      &   & + 4\, (1-\cos\alpha_1)   (1-\cos\alpha_2) \, \lambda^2 \}
      \sqrt{1-\lambda}.
\end{eqnarray} 
The factor multiplying $\sqrt{1-\lambda}$ is quadratic in $\lambda$
and hence it can vanish for two values of $\lambda$.  Thus we can ask
ourselves the questions, suppose two values of $\lambda$ between zero
and one are given at which $P(\lambda)=1$, what are the corresponding
values of $\alpha$ and what limits are there on the possible values of
$\lambda$ that satisfy $P(\lambda)=1$?  If $\lambda_1$ and $\lambda_2$
are the roots of the equation
\begin{equation}\label{eq:17a}
  u_2(\lambda)/\sqrt{1-\lambda}=0,
\end{equation}
then $\cos\alpha_1$ and $\cos\alpha_2$ satisfy the equations
\begin{widetext}
\begin{eqnarray}\label{eq:18}
&&  8\lambda_1\lambda_2\cos^3\alpha_2 + [4(\lambda_1+\lambda_2)
  (1-\lambda_1-\lambda_2) - 8\lambda_1\lambda_2]\cos^2\alpha_2
  + [8(\lambda_1+\lambda_2)^2 - 12(\lambda_1+\lambda_2)-8\lambda_1\lambda_2
  + 4]\cos\alpha_2 \nonumber \\
&& \hspace{1.7in} - 4(\lambda_1+\lambda_2)^2 + 8(\lambda_1+\lambda_2) 
  -5 + 8\lambda_1\lambda_2 = 0,
\end{eqnarray}
\begin{equation}\label{eq:19}
  \cos\alpha_1 = 1 - \frac{1}{4(1-\cos\alpha_2)\lambda_1\lambda_2}. 
\end{equation}                                                                            
\end{widetext}
In order to have a sense of the values of $\alpha_1$ and $\alpha_2$
that are valid, we have minimally the condition that the discriminant
of Eq.~(\ref{eq:17a}) (quadratic in $\lambda$) should be nonnegative
to avoid complex values of $\lambda$.  In Fig.~{\ref{fig:3a}} we plot
the discriminant as a surface $z= D(\alpha_1,\alpha_2)$; the
intersection of the surface with the $xy$ plane gives the boundary of
the non-allowed $\alpha_1$ and $\alpha_2$ values.

Given $\lambda_1$ and $\lambda_2$ one can solve Eq.~(\ref{eq:18}) for
$\cos\alpha_2$ and using it we obtain $\cos\alpha_1$ from the second
equation.  Only those solutions that yield real angles $\alpha_1$ and
$\alpha_2$ are meaningful for the unitary operators.  The minimum
value of $\lambda$ for which $P=1$ occurs when $\alpha_1=\alpha_2 =
\pi$.  In that case $\lambda= (3-\sqrt{5})/8 =0.09549$.  It can be
shown that varying $\alpha_1$ or $\alpha_2$ by a small amount away from
$\pi$ always leads to an increase in the $\lambda$ which corresponds
to the smaller of the two values of $\lambda$.  When we let
$\alpha_{1,2}=\pi+\epsilon_{1,2}$ we obtain a change in the smaller
$\lambda$ of
\begin{equation}\label{eq:20}
  \Delta\lambda = \frac{1}{160}\left[\left(2\sqrt{5}\epsilon_1
  +\frac{5-3\sqrt{5}}{\sqrt{2\sqrt{5}}}\epsilon_2\right)^2+(22-8\sqrt{5})
\epsilon_2^2\right],
\end{equation}
which is positive regardless of the signs of $\epsilon_{1,2}$.  The
larger $\lambda$ can increase or decrease with changes in the
phases(s).

\begin{figure}[h]
  \centering
   \resizebox{3.5in}{!}{\includegraphics{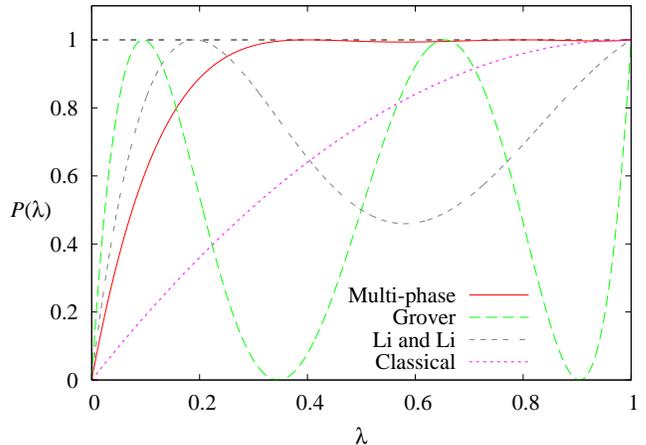}}
   \caption{(Color online) Plots of $P(\lambda)$ obtained after two
     iterations of the multiphase case with $\alpha_1=1.00889485$ and
     $\alpha_2=2.30794928$.  For comparison the double Grover
     iteration ($\alpha_1=\alpha_2=\pi$), the double iteration of Li
     and Li ($\alpha_1=\alpha_2=\pi/2$), and the classical counterpart
     are shown as well.}
\label{fig:3}
\end{figure}

We obtain a particular example using the procedure described above.
We search through combinations of $\lambda_{1,2}$ and find that
$\lambda_1 = 2/5$ and $\lambda_2=4/5$ give good results.  In this case
$\alpha_1=1.00889485$ and $\alpha_2=2.30794928$.  We find local minima
of $P(\lambda)$ at $\lambda = 0.5767$ and $\lambda = 0.9433$ at which
$P = 0.9936$ and 0.9966, respectively.  The corresponding graph of the
success probability as a function of $\lambda$ obtained with the
two-stage multi-phase operator is shown in Fig.~\ref{fig:3} and
compared with double iterations of the Grover operation and that of Li
and Li~\cite{li07}.

It would be interesting to examine a classical counterpart of
$P(\lambda)$.  The probability of failing to find one of $M$ marked
objects out of $N$ objects is $(N-M)/N = 1-\lambda$.  The probability
of failing twice in a row is
  $$ (1-\lambda)\left(\frac{N-1-M}{N-1}\right)=(1-\lambda)
  \left(1-\frac{\lambda}{1-1/N}\right). 
  $$
  The probability of failing $k$ times in a row is
  \begin{eqnarray}
  && (1-\lambda)\left(1-\frac{\lambda}{1-1/N}\right)\cdots
  \left(1-\frac{\lambda}{1-(k-1)/N}\right) \nonumber \\
  &&=\prod_{n=1}^k
    \left[1-\lambda\left(1-\frac{n-1}{N}\right)^{-1}\right].
  \nonumber
\end{eqnarray}
Thus the probability of finding at least one of the $M$ items in $k$
successive attempts is
\begin{equation}\label{eq:class}
  P_\mathrm{classical}(\lambda) = 1 - 
  \prod_{n=1}^k\left[1-\lambda\left(1-\frac{n-1}{N}\right)^{-1}\right].
\end{equation}
If $k\ll N$, this probability is approximately
$P_\mathrm{classical}(\lambda)\approx 1-(1-\lambda)^k$, which we
interpret as the classical counterpart of $P(\lambda)$.  This
probability with $k=2$ is also plotted in Fig.~\ref{fig:3}.

\subsection{Multi-phase matching with six pairs of phases}
\label{six}

We show that if we match the multi-phase $\alpha_j$ and $\beta_j$
$(j=1,\dots,k)$ with $k=6$ in accordance with a certain matching rule
(best fit), we can obtain a multiple Grover operation that yields
$P(\lambda)\approx 1$ in a wide range of $\lambda$.  We found this
best solution for six Grover iterations by a nonlinear fitting to the
ideal probability curve $P(\lambda)=1$ for $0<\lambda\leq 1$. The
phases $\alpha_j$ and $\beta_j$ found in this way are given in the
left side of Table~\ref{table:1}.
\begin{table}[ht]
\begin{tabular}{clc|cc}
\hline
$~~j~~$ & \multicolumn{1}{c}{$\alpha_j/\pi$} & $\beta_j$ & 
  $\lambda_j^{(P(\lambda_j)=1)}$ & $P(\lambda_j^{(\mathrm{local~min})})$ \\
\hline\hline
1 & 1.20560132~ & ~$-\alpha_6$~ & 0.10777 & 0.9980 \\
2 & 1.29806396 & $-\alpha_5$ & 0.23793 & 0.9993 \\
3 & 1.31701508 & $-\alpha_4$ & 0.41889 & 0.9996 \\
4 & 1.33356767 & $-\alpha_3$ & 0.62393 & 0.9997 \\
5 & 0.47289426 & $-\alpha_2$ & 0.81366 & 0.9997 \\
6 & 1.66668634 & $-\alpha_1$ & 0.94483 & 0.9995 \\
\hline
\end{tabular}
\caption{Phase parameters for the six-parameter multi-phase matching, and 
  results for local maxima ($P(\lambda)=1$) and 
  local minima of $P(\lambda)$.}
\label{table:1}
\end{table}
  
\begin{figure}[h]
  \centering
   \resizebox{3.5in}{!}{\includegraphics{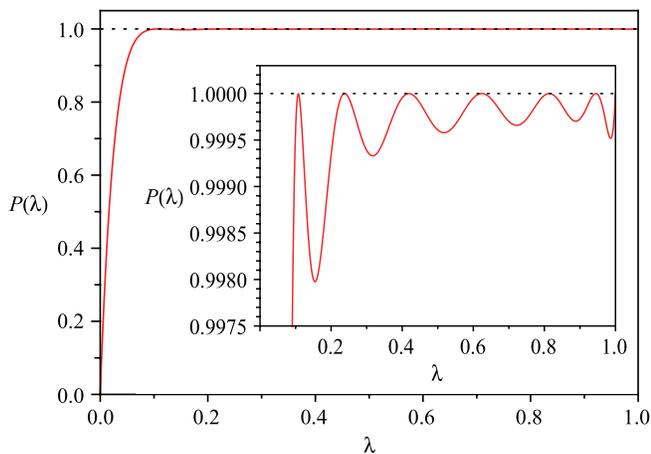}}
   \caption{(Color online) Success probability curves $P(\lambda)$
     obtained with the multi-phase matching with six iteration in the
     operation.  The phase parameters are given in the left side of
     Table~\ref{table:1}.  }
\label{fig:4}
\end{figure}

It is remarkable that $\alpha_j$ and $\beta_j$ are matched to each
other such that $\alpha_j=-\beta_{6-j+1}$.  The signs of $\alpha_j$
and $\beta_j$ are opposite to each other, which is consistent with the
case of the new phase matching of Ref.~\cite{li07}, i.e., the $k=1$
case with $\alpha_1=-\beta_1 = \pi/2$.  The matching rule $\alpha_j =
-\beta_{k-j+1}$ between the multi-phases $\alpha_j$ and $\beta_j$
holds for any $k$ in the best solution obtained by the nonlinear
fitting to the ideal probability curve $P(\lambda)=1$ for $0 < \lambda
\leq 1$, although we omit to show cases other than those for which
$k=1$, 2, and 6.

Fig.~\ref{fig:4} shows the success probabilities obtained by six
Grover operations with the multi-phase matching of Eq.~(\ref{eq:7}).
The inset of Fig.~\ref{fig:4} shows that there are six values of
$\lambda_i$ at which $P(\lambda_i)=1$ exactly.  They are 
given in the right side of Table~\ref{table:1} along with local
minimum values of the function $P(\lambda)$ which occur between 0.1
and 1.  

We studied the $k=5$ case in the same way and obtained a graph similar
to Fig.~\ref{fig:4} with $P(\lambda)=1$ for \emph{five} values of
$\lambda$ other than unity. The local minima of $P(\lambda)$ are lower
and the minimum value of $\lambda$ for which $P(\lambda)=1$ is
slightly larger than in the $k=6$ case.  The matching rule
$\alpha_j=-\beta_{k-j+1}$ is also satisfied for the $k=5$ case as it
was for $k=1$, 2, and 6 cases.  We are confident that for any $k>1$
this matching rule for the best fit holds so that in general one finds
$k$ values of $\lambda$ for which $P(\lambda)=1$ and the smallest
$\lambda$ for which $P(\lambda)=1$ decreases as $k$ increases.

\begin{figure}[h]
  \centering
   \resizebox{3.5in}{!}{\includegraphics{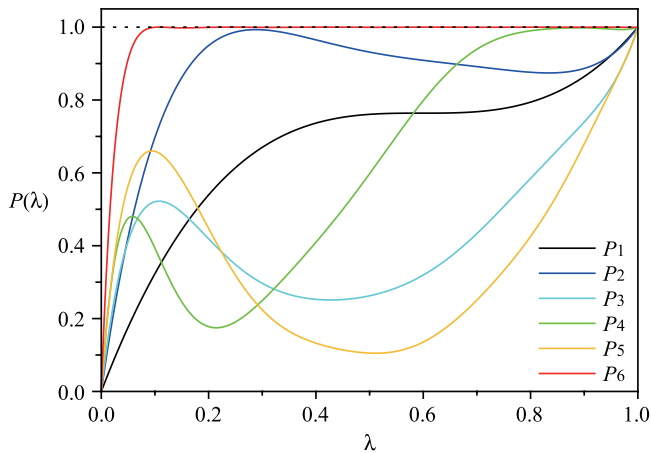}}
   \caption{(Color online) Success probability curves $P_i(\lambda)$
     at the $i$th stage obtained with the six-stage multi-phase
     matching.  The phase parameters are given in the left side of
     Table~\ref{table:1}.  }
\label{fig:5}
\end{figure}

Returning to the six-stage multiple phase operation, we define
$P_j(\lambda)$ $(j=1,\dots,6)$ as the success probability curves after
$j$ steps of the six-stage multi-phase operation.  As seen in the
Figs.~\ref{fig:4} and \ref{fig:5}, $P_6(\lambda)\approx 1$ is achieved
for $0.1\leq\lambda\leq 1$ in the sixth Grover operation.  This is
significant in the sense that if $\lambda$ is greater than 0.1, we can
always find the superposition of marked states by just six Grover
operations.  In contrast to the shape of the curve for $P_6(\lambda)$,
Fig.~\ref{fig:5} shows that each $P_j(\lambda)$ for $j=1,\dots,5$
depends strongly on $\lambda$ and is far from the desired success
probability $P_6(\lambda)$.  The curves do not monotonically approach
the desired success probability $P_6(\lambda)$ when $\lambda > 0.05$.
In particular, $P_5(\lambda)$ is quite different from the desired
probability $P_6(\lambda)$. However, in the final (sixth) step the
desired probability $P_6(\lambda)$ is obtained.  This is in contrast
to the fixed-point iteration schemes studied in
Refs.~\cite{grover05,tulsi06,younes07}.

\begin{figure}[ht]
  \centering \resizebox{3.5in}{!}{\includegraphics{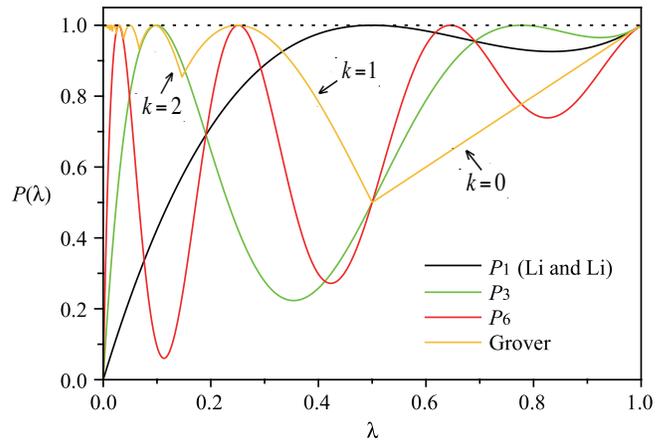}}
  \caption{(Color online) Success probability curves $P_1(\lambda)$,
    $P_3(\lambda)$, and $P_6(\lambda)$ obtained with the single phase
    matching with $\alpha_j=-\beta_j=\pi/2$ $ (j=1, \cdots, 6)$. The
    $P_1(\lambda)$ indicates the success probability of the new phase
    matching of Ref.~\cite{li07}. The yellow curve is the success
    probability of the original Grover's algorithm, where the graph
    segments are plotted for optimal iteration times indicated by $k$.
    Iteration times for small $\lambda$ are omitted.}
\label{fig:6}
\end{figure}

Figure~\ref{fig:6} shows the success probabilities obtained by six
Grover operations with the single phase matching with
$\alpha_j=-\beta_j=\pi/2$ $(j=1,\dots,6)$ , where we showed only
$P_1(\lambda)$, $P_3(\lambda)$ and $P_6(\lambda)$.  The $P_1(\lambda)$
is the success probability of the new phase matching obtained by Li
and Li~\cite{li07}. As stressed in Ref.~\cite{li07}, the success
probability is substantially improved in $\lambda>1/3$ by a single
Grover operation, compared with that of the original Grover algorithm
indicated by the yellow line, where the probability is plotted for
optimal iteration times indicated by $k$.  However, $P_3(\lambda)$ and
$P_6(\lambda)$ obtained by multiple Grover operations with the single
phase matching show intensive oscillations with $\lambda$. As we have
shown, such undesirable oscillations can be eliminated by the
multi-phase matching subject to the matching rule
$\alpha_j=-\beta_{6-j+1}$.

Here we should note that the nonlinear fitting is not unique.  The
phases $\alpha_j$ and $\beta_j$ given in Table~\ref{table:1} were
obtained by minimizing the function $\sum_i\chi_i^2$ where the
$\chi_i$ are the differences at $\lambda=\lambda_i$ of the ideal
probability $P(\lambda)=1$ and the probability function
$P_6(\lambda)$.  If we take, for example, a function such as
$\sum_i|\chi_i|$ we obtain another solution.  Although this solution
gives almost the same $P_6(\lambda)$, the probability curve is shifted
slightly toward larger values of $\lambda$, so that the local extrema
are also slightly moved to the right.  Since we emphasize obtaining
$P(\lambda)\approx 1$ over as wide a range of $\lambda$ as possible we
adopted the solution that uses the $\chi^2_i$ for the fitting.

\section{Iteration of Grover's operation with phase other than $\pi$}
\label{k_iterations}

In this section we consider the repeated application of Grover's
original operation generalized to have a phase other than $\pi$.  We
focus in particular on cases with small $\lambda$ for which the
success probability with the multi-phase matching is small, and
determine the conditions that yield success probabilities close to
unity.

\begin{figure}[h]
  \centering
   \resizebox{3.5in}{!}{\includegraphics{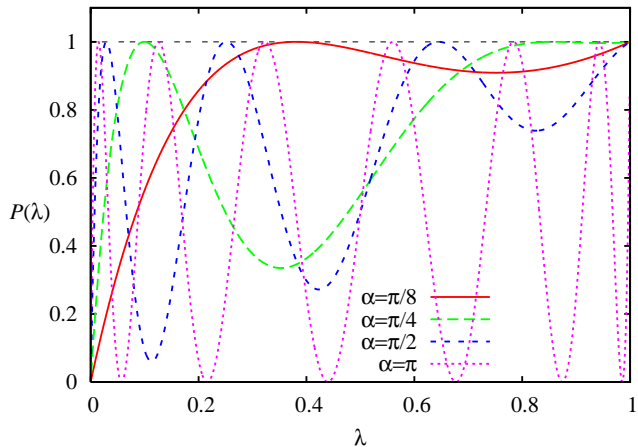}}
   \caption{(Color online) Plots of $P(\lambda)$ obtained after six
     iterations with a single $\alpha$ for the cases of $\alpha =
     \pi/8, \ \pi/4, \ \pi/2$, and $\pi$.}
\label{fig:ap1}
\end{figure}

Consider a single Grover operation with matched phase,
Eq.~(\ref{eq:8}), but with $\beta=-\alpha$,
\begin{equation}\label{eq:a1}
  G_1 = \begin{pmatrix} (1-e^{-i\alpha})(1-\lambda)+ e^{-i\alpha} & 
    (e^{i\alpha}-1)\sqrt{\lambda(1-\lambda)} \\
    (1-e^{-i\alpha})\sqrt{\lambda(1-\lambda)} &
    (e^{i\alpha}-1)\lambda+1
    \end{pmatrix}.
\end{equation}
Note that $\det G_1 = 1$.  We obtain eigenvalues $\sigma$ of the
matrix $G_1$ by solving
\begin{equation}\label{eq:a2}
  f(\sigma)=\det(G_1-\sigma I) = 0.
\end{equation}  
The characteristic function $f(\sigma)$ is
\begin{equation}\label{eq:a3}
  f(\sigma) = \sigma^2 + 2[-1+(1-\cos\alpha)\lambda]\sigma + 1.
\end{equation}
The equation $f(\sigma)=0$ yields solutions
\begin{equation}\label{eq:a4}
  \sigma = 1 - (1-\cos\alpha)\lambda \pm i\sqrt{(1-\cos\alpha)\lambda
    [2-(1-\cos\alpha)\lambda]}
\end{equation}
We define $x$ as
\begin{equation}\label{eq:a4a}
 x =(1-\cos\alpha)\lambda,
\end{equation} 
so that the eigenvalues can be written as
\begin{equation}\label{eq:a5}
  \sigma = e^{\pm i\phi}, \ \ \phi = 
\arctan\left(\frac{\sqrt{x(2-x)}}{1-x}\right).
\end{equation}
We choose the definition of the arc tangent so that as $x$ varies from
0 to 2, $\phi$ goes from 0 to $\pi$.
We can rewrite the function $f(\sigma)$ as
\begin{equation}\label{eq:a6}
  f(\sigma) = \sigma^2 - 2\sigma\cos\phi +1.
\end{equation}
By the Cayley-Hamilton theorem~\cite[page 91]{pipes58} $f(G_1) =0$, so
that we obtain the identity
\begin{equation}\label{eq:a7}
  G_1^2 = 2G_1\cos\phi -1.
\end{equation}
This means that $G_1$ iterated any number of times can be written as a
linear expression of $G_1$.  In fact for $k$ iterations it can be
shown by induction~\cite{sprung93} that
\begin{equation}\label{eq:a8}
  G_1^k = \frac{1}{\sin\phi}\left[G_1\sin(k\phi) - \sin((k-1)\phi)\right].
\end{equation}
Consider now the $k$ iterations of the Grover operation, so that
\begin{equation}\label{eq:a9}
  \begin{pmatrix} u_k \\ d_k \end{pmatrix} = G_1^k
  \begin{pmatrix} \sqrt{1-\lambda} \\ \sqrt{\lambda} \end{pmatrix}.
\end{equation}
This yields
\begin{widetext}
\begin{equation}\label{eq:a10}
 \begin{pmatrix} u_k \\ d_k \end{pmatrix} = \frac{1}{\sin\phi} \left[
\sin{(k\phi)} \begin{pmatrix} (1-e^{-i\alpha})(1-\lambda)+ e^{-i\alpha} & 
    (e^{i\alpha}-1)\sqrt{\lambda(1-\lambda)} \\
    (1-e^{-i\alpha})\sqrt{\lambda(1-\lambda)} &
    (e^{i\alpha}-1)\lambda+1
    \end{pmatrix} - \sin{((k-1)\phi)} \ I \right]
\begin{pmatrix} \sqrt{1-\lambda} \\ \sqrt{\lambda} \end{pmatrix}.
\end{equation}
\end{widetext}
Thus the expression for $u_k$ is
\begin{equation}\label{eq:a11}
  u_k = \frac{\sqrt{1-\lambda}}{\sin\phi}\left\{\sin{(k\phi)}(1-2x)
    -\sin{((k-1)\phi)}
  \right\}.
\end{equation}
We require $u_k=0$ so that $P=1$.  A trivial solution is $\lambda=1$.
We also note that $\phi=0$ yields $x=-1/(2k)$.  Since $x$ must be
positive $\sin{\phi}\neq 0$.  Thus we need to solve only
\begin{equation}\label{eq:a12}
  \sin{(k\phi)}(1-2x)-\sin{((k-1)\phi)} = 0.
\end{equation}
The solutions are values of $x=(1-\cos{\alpha})\lambda$ for which
$P=1$.  Thus we have $P=1$ for combinations of $\alpha$ and $\lambda$.
For instance, when $\alpha=\pi$, then $\lambda = x/2$.  In
Fig.~\ref{fig:ap1}, we display the $P(\lambda)$ curves for six ($k=6$)
iterations when $\alpha$ has different values.

For large $k$ we can estimate the smallest value of $\lambda$ for which
$P$ is unity.  We rewrite Eq.~(\ref{eq:a12}) so that
\begin{equation}\label{eq:a13}
  \tan(k\phi) = \frac{\sin\phi}{\cos\phi-1+2x}.
\end{equation}
The value of $x$ which is the solution occurs for the $x$ coordinate
of the point of intersection of the curves represented by the left
side and the right side of Eq.~(\ref{eq:a13}).  The curve on the right
is a smoothly decreasing positive function starting at infinity when
$x=0$ and asymptotically approaching the positive $x$ axis.  The curve
on the left starts at zero and increases to positive infinity when
$k\phi= \pi/2$.  When $k$ is large this occurs for small values of
$\phi$ or small values of $x$.  Thus using the condition $\phi\approx 
\pi/(2k)$, we obtain
\begin{equation}\label{eq:a14}
  \arctan\frac{\sqrt{x(2-x)}}{1-x} \lapproxeq \frac{\pi}{2k} 
  \ \ \mathrm{or} \ \  \frac{\sqrt{x(2-x)}}{1-x} \lapproxeq \frac{\pi}{2k}.
\end{equation}
This leads to the approximation of the smallest value of $x$ for which
$P$ is one as $x_\mathrm{min} \lapproxeq \pi^2/(8k^2)$; for $\alpha =
\pi$ (the Grover case) $\lambda_\mathrm{min}= x_\mathrm{min}/2$.  This
approximation leads to $\lambda_\mathrm{min} = 0.017$ for $k=6$ and
$\alpha=\pi$, whereas the exact solution of Eq.~(\ref{eq:a13}) gives
$0.014$.  As $k$ gets larger the approximation improves further.

\subsection{The Grover algorithm as a special case}
We recover the Grover algorithm starting with Eq.~(\ref{eq:a11}) and
setting $\alpha=\pi$ or $x=2\lambda$.  Then it follows from
Eq.~(\ref{eq:a5}) that $\sin\phi = \sqrt{4\lambda(1-\lambda)}$ and
$\cos\phi = (1-2\lambda)$.  The $u_k$ of Eq.~(\ref{eq:a11}) can be
reduced to
\begin{equation}\label{eq:a15}
  u_k=-\sqrt{\lambda}\sin(k\phi) + \sqrt{1-\lambda}\cos(k\phi).
\end{equation}
Define $\sin\theta = \sqrt{\lambda}$.  Then
\begin{equation}\label{eq:a16}
  u_k=\cos(k\phi+\theta).
\end{equation}
We can show that $\phi = 2\theta$, so that
\begin{equation}\label{eq:a17}
  u_k = \cos[(2k+1)\theta] = \cos[(2k+1)\arcsin(\sqrt{\lambda})].
\end{equation}
This is Eq.~(6) of Ref.~\cite{li07}.  Furthermore
\begin{equation}\label{eq:a18}
  P=1-u_k^2 = \sin^2[(2k+1)\arcsin(\sqrt{\lambda})].
\end{equation}
Thus each iteration effectively rotates the state through an angle of
$\theta/2 = \arcsin(\sqrt{\lambda})/2$.  We can use this to estimate
the number of iterations that are required to obtain $P(\lambda) =1$.
That occurs when the argument of the sine function in
Eq.~(\ref{eq:a18}) is $\pi/2$, i.e.,
\begin{equation}\label{eq:a19}
  k=\frac{1}{2}\left(\frac{\pi}{2\theta}-1\right).
\end{equation}
For small $\theta$ (or large $k$) 
\begin{equation}\label{eq:a20}
  k \approx\frac{\pi}{4\theta} \approx \left[\frac{\pi}{4\theta}\right] 
= \mathrm{~integer~value~of~} \frac{\pi}{4\theta} \approx \left[
\frac{\pi}{4}\frac{1}{\sqrt{\lambda}}\right].
\end{equation}
Thus after approximately $\pi/(4\sqrt{\lambda})$ iterations one has
certainty of having found the superposition of marked states.
Classically the number of search operations to have this certainty is
on the average approximately $1/(2\lambda)=N/(2M)$ for $N$ much larger
than $M$.  By the same reasoning we find $P=0$ with twice as many
quantum iterations.  Thus by continuing to iterate indefinitely we can
end with any probability of success.  However, if we iterate close to
the number that gives 100\% probability of success we have a good
approximation to a successful search.

\subsection{Effect of phase  $\mathbf{\alpha}$}
\label{general_phase}
\begin{figure}[h]
  \centering
   \resizebox{3.5in}{!}{\includegraphics{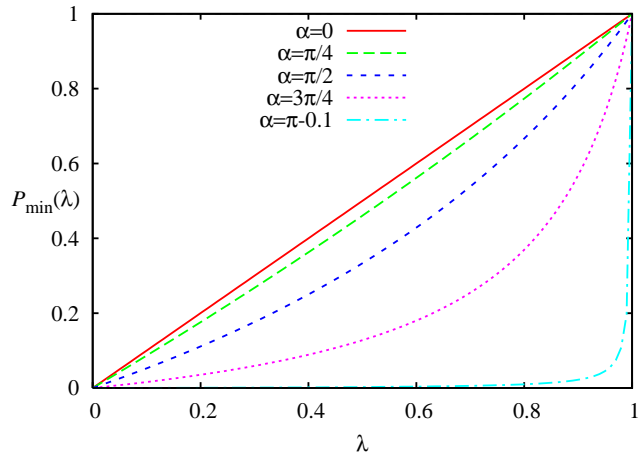}}
   \caption{(Color online) Plots of $P_\mathrm{min}(\lambda)$ for 
     $\alpha = 0, \ \pi/4, \ \pi/2, \ 3\pi/4$, and $\pi-0.1$.}
\label{fig:ap2}
\end{figure}

For a general value of $x=(1-\cos\alpha)\lambda$, Eq.~(\ref{eq:a11}) can
be written as
\begin{equation}\label{eq:a21}
u_k = A\cos(k\phi+\theta),
\end{equation}
where
\begin{equation}\label{eq:a22}
A=\sqrt{\frac{2(1-\lambda)}{2-x}}, \ \ \ \theta = 
\arctan\sqrt{\frac{x}{2-x}}.
\end{equation}
Since $P(\lambda) = 1 - u_k^2 = 1 - A^2\cos^2(k\phi+\theta)$, the minimum
of $P(\lambda)$ is
\begin{equation}\label{eq:a23}
  P_\mathrm{min}(\lambda) = 1 - A^2 = \frac{\lambda(1+\cos\alpha)}
  {2-(1-\cos\alpha)\lambda}.
\end{equation}
In Fig.~\ref{fig:ap2} it is seen that $P_\mathrm{min}(\lambda)$ has at
most a linear rise as $\lambda$ increases from zero to one.

\section{Summary}
\label{summary}
We have proposed a multi-phase matching for the Grover search
algorithm, which is an extension of the new phase matching proposed in
Ref.~\cite{li07}. The multi-phase matching is characterized by
multiple Grover operations with two kinds of multi-phases $\alpha_j$
and $\beta_j$ $(j=1,\dots,k)$. We showed that if we match $\alpha_j$
and $\beta_j$ in accordance with the rule $\alpha_j = -\beta_{k-j+1}$
for a given $k$ we can obtain an optimal solution for $\alpha_j$,
$\beta_j$ that gives a success probability curve such that it is
almost constant and unity in a wide range of the fraction of marked
states.  As an example we presented an optimal solution obtained for
$k=6$. The solution yields the desired success probability $P = 1$ to
within 0.2\% for the fraction of the marked states greater than 0.1.
This is significant in the sense that when the fraction of marked
states is greater than 0.1, we can always with a high degree of
confidence find a uniform superposition of the marked states by
repeating the Grover operation just six times.

To clarify the mechanism of the multi-phase matching we studied in
detail the one- and two-iteration cases.  We showed that it is
possible to obtain $P=1$ exactly for a particular fraction $\lambda$
by tuning the phases.  This can be generalized to having $k$ values of
$\lambda$ for which $P(\lambda)=1$ when we go to a $k$-iteration
scheme.

One can obtain $P=1$ for a given very small $\lambda$ by using the
original Grover algorithm or the phase-matched version of it.  In this
case usually a specified large number of iterations is required.
Further study is needed to obtain an efficient algorithm for extremely
small $\lambda$.

\begin{acknowledgments}
 This work was supported by the Japan Society for the
Promotion of Sciences and the Natural Sciences and Engineering
Research Council of Canada.
\end{acknowledgments}

\appendix

\section{Equivalence of two phase-matching schemes}

Li and Li~\cite{li07} claim to have generalized the Long
phase-matching algorithm in order to produce a higher success
probability.  In actual fact the phase matching of Li and Li and that
of Long \emph{et al.}~\cite{long99} result in the same success probability.
We show that in the following.  

Instead of Eqs.~(\ref{eq:1}) and (\ref{eq:2}), Long \emph{et al.} work
with the operators
\begin{equation}\label{eq:aa1}
U=I-(1-e^{i\theta})\sum_{l=0}^{M-1}|t_l\rangle\langle t_l|  
\end{equation} 
\begin{equation}\label{eq:aa2}
  V=I-(1-e^{i\phi})|0^{\otimes n}\rangle\langle 0^{\otimes n}|.  
\end{equation} 
These unitary transformations lead to the Grover operator (in the
notation of this paper) $G(\theta,\phi)$, where
\begin{equation}\label{eq:aa3}
  G = 
\begin{pmatrix}
      1-(1-e^{i\phi})(1-\lambda)  &
      -(1 - e^{i\phi})e^{i\theta}\sqrt{\lambda(1-\lambda)} \\
      -(1-e^{i\phi})\sqrt{\lambda(1-\lambda)} &
      [1-(1 - e^{i\phi})\lambda]e^{i\theta}
      \end{pmatrix}. 
\end{equation}
For one operation we calculate the final state
\begin{equation}\label{eq:aa4}
  \begin{pmatrix}u \\ d \end{pmatrix} =  G(\theta,\phi)
  \begin{pmatrix}\sqrt{1-\lambda} \\ \sqrt{\lambda} \end{pmatrix}
\end{equation}
with
\begin{equation}\label{eq:aa5}
  u=\sqrt{1-\lambda}[1-(1-e^{i\phi})(1-\lambda) - (1-e^{i\phi})
  e^{i\theta}\lambda].
\end{equation}
Setting $u=0$ we obtain (in addition to $\lambda=1$) the solution 
\begin{equation}\label{eq:aa6}
  \phi=\theta, \ \ \ \lambda= \frac{1}{2}\frac{\cos\theta +1}{\sin^2\theta}. 
\end{equation}
Note that the signs of $\phi$ and $\theta$ are the same, unlike the
opposite signs of the matched phases of Li and Li, i.e.,
$\beta=-\alpha$.  In order that $0 < \lambda\leq 1$ with this phase
matching, $\theta$ varies from $\pi/3$ to $\pi$.  For $P(\lambda) =
1-|u|^2$, we obtain the expression of Eq.~(\ref{eq:13}) with $\alpha$
replaced by $\theta$.  Thus the impressive result by Li and Li of a
single phase-matched Grover operation can also be obtained with the
earlier-proposed operation of Long \emph{et al.}  However, the
formulation of Li and Li results in $\mathrm{Im} \, u = 0$ when
$\beta = -\alpha$, whereas $\mathrm{Im} \, u \neq 0$ when
$\phi=\theta$ for the operator of Long \emph{et al.}  It should be
noted however that the remarkable single-operation result was first
reported by Li and Li~\cite{li07}.  Although the probabilities are the
same the amplitudes are not, and Li and Li's formulation gives a more
straightforward derivation of the probabilities. (See Sec. IIA.)  One
can relate the two formulations by suggesting that instead of the
operator acting on $(\sqrt{1-\lambda},\sqrt{\lambda})^T$
initially, in the case of Long \emph{et al.} it operates on this
state multiplied by a phase factor.


\begin{thebibliography}{17}

\expandafter\ifx\csname natexlab\endcsname\relax\def\natexlab#1{#1}\fi
\expandafter\ifx\csname bibnamefont\endcsname\relax
  \def\bibnamefont#1{#1}\fi
\expandafter\ifx\csname bibfnamefont\endcsname\relax
  \def\bibfnamefont#1{#1}\fi
\expandafter\ifx\csname citenamefont\endcsname\relax
  \def\citenamefont#1{#1}\fi
\expandafter\ifx\csname url\endcsname\relax
  \def\url#1{\texttt{#1}}\fi
\expandafter\ifx\csname urlprefix\endcsname\relax\def\urlprefix{URL }\fi
\providecommand{\bibinfo}[2]{#2}
\providecommand{\eprint}[2][]{\url{#2}}

\bibitem[{\citenamefont{Grover}(1996)}]{grover96}
\bibinfo{author}{\bibfnamefont{L.~K.} \bibnamefont{Grover}}, in
  \emph{\bibinfo{booktitle}{STOC '96: Proceedings of the twenty-eighth annual
  ACM symposium on theory of computing}} (\bibinfo{publisher}{ACM},
  \bibinfo{address}{New York, NY, USA}, \bibinfo{year}{1996}), pp.
  \bibinfo{pages}{212--219}.

\bibitem[{\citenamefont{Grover}(1997{\natexlab{a}})}]{grover97a}
\bibinfo{author}{\bibfnamefont{L.~K.} \bibnamefont{Grover}},
  \bibinfo{journal}{Phys. Rev. Lett.} \textbf{\bibinfo{volume}{79}},
  \bibinfo{pages}{325} (\bibinfo{year}{1997}{\natexlab{a}}).

\bibitem[{\citenamefont{Grover}(1997{\natexlab{b}})}]{grover97b}
\bibinfo{author}{\bibfnamefont{L.~K.} \bibnamefont{Grover}},
  \bibinfo{journal}{Phys. Rev. Lett.} \textbf{\bibinfo{volume}{79}},
  \bibinfo{pages}{4709} (\bibinfo{year}{1997}{\natexlab{b}}).

\bibitem[{\citenamefont{Grover}(2001)}]{grover01}
\bibinfo{author}{\bibfnamefont{L.~K.} \bibnamefont{Grover}},
  \bibinfo{journal}{Am. J. Phys.} \textbf{\bibinfo{volume}{69}},
  \bibinfo{pages}{769} (\bibinfo{year}{2001}).

\bibitem[{\citenamefont{Grover}(1998)}]{grover98}
\bibinfo{author}{\bibfnamefont{L.~K.} \bibnamefont{Grover}},
  \bibinfo{journal}{Phys. Rev. Lett.} \textbf{\bibinfo{volume}{80}},
  \bibinfo{pages}{4329} (\bibinfo{year}{1998}).

\bibitem[{\citenamefont{Long et~al.}(1999)\citenamefont{Long, Li, Zhang, and
  Niu}}]{long99}
\bibinfo{author}{\bibfnamefont{G.~L.} \bibnamefont{Long}},
  \bibinfo{author}{\bibfnamefont{Y.~S.} \bibnamefont{Li}},
  \bibinfo{author}{\bibfnamefont{W.~L.} \bibnamefont{Zhang}}, \bibnamefont{and}
  \bibinfo{author}{\bibfnamefont{L.}~\bibnamefont{Niu}},
  \bibinfo{journal}{Phys. Lett. A} \textbf{\bibinfo{volume}{262}},
  \bibinfo{pages}{27} (\bibinfo{year}{1999}).

\bibitem[{\citenamefont{Long et~al.}(2001)\citenamefont{Long, Yan, Li, Tu, Tao,
  Chen, Liu, Zhang, Luo, Xiao et~al.}}]{long01}
\bibinfo{author}{\bibfnamefont{G.~L.} \bibnamefont{Long}},
  \bibinfo{author}{\bibfnamefont{H.}~\bibnamefont{Yan}},
  \bibinfo{author}{\bibfnamefont{Y.~S.} \bibnamefont{Li}},
  \bibinfo{author}{\bibfnamefont{C.~C.} \bibnamefont{Tu}},
  \bibinfo{author}{\bibfnamefont{J.~X.} \bibnamefont{Tao}},
  \bibinfo{author}{\bibfnamefont{H.~M.} \bibnamefont{Chen}},
  \bibinfo{author}{\bibfnamefont{M.~L.} \bibnamefont{Liu}},
  \bibinfo{author}{\bibfnamefont{X.}~\bibnamefont{Zhang}},
  \bibinfo{author}{\bibfnamefont{J.}~\bibnamefont{Luo}},
  \bibinfo{author}{\bibfnamefont{L.}~\bibnamefont{Xiao}}, \bibnamefont{et~al.},
  \bibinfo{journal}{Phys. Lett. A} \textbf{\bibinfo{volume}{286}},
  \bibinfo{pages}{121} (\bibinfo{year}{2001}).

\bibitem[{\citenamefont{Collins}(2002)}]{collins02}
\bibinfo{author}{\bibfnamefont{D.}~\bibnamefont{Collins}},
  \bibinfo{journal}{Phys. Rev. A} \textbf{\bibinfo{volume}{65}},
  \bibinfo{pages}{052321} (\bibinfo{year}{2002}).

\bibitem[{\citenamefont{Tulsi et~al.}(2006)\citenamefont{Tulsi, Grover, and
  Patel}}]{tulsi06}
\bibinfo{author}{\bibfnamefont{T.}~\bibnamefont{Tulsi}},
  \bibinfo{author}{\bibfnamefont{L.~K.} \bibnamefont{Grover}},
  \bibnamefont{and} \bibinfo{author}{\bibfnamefont{A.}~\bibnamefont{Patel}},
  \bibinfo{journal}{Quant. Inform. Comp.} \textbf{\bibinfo{volume}{6}},
  \bibinfo{pages}{483} (\bibinfo{year}{2006}).

\bibitem[{\citenamefont{Li and Li}(2007)}]{li07}
\bibinfo{author}{\bibfnamefont{P.}~\bibnamefont{Li}} \bibnamefont{and}
  \bibinfo{author}{\bibfnamefont{S.}~\bibnamefont{Li}}, \bibinfo{journal}{Phys.
  Lett. A} \textbf{\bibinfo{volume}{366}}, \bibinfo{pages}{42}
  (\bibinfo{year}{2007}).

\bibitem[{\citenamefont{Younes}(2007)}]{younes07}
\bibinfo{author}{\bibfnamefont{A.}~\bibnamefont{Younes}},
  \bibinfo{journal}{arXiv:quant-ph/0704.1585v2}  (\bibinfo{year}{2007}).

\bibitem[{\citenamefont{Chi and Kim}(1997)}]{chi97}
\bibinfo{author}{\bibfnamefont{D.~P.} \bibnamefont{Chi}} \bibnamefont{and}
  \bibinfo{author}{\bibfnamefont{J.}~\bibnamefont{Kim}},
  \bibinfo{journal}{arXiv:quantum-ph/9708005v1}  (\bibinfo{year}{1997}).

\bibitem[{\citenamefont{H\o{}yer}(2000)}]{hoyer00}
\bibinfo{author}{\bibfnamefont{P.}~\bibnamefont{H\o{}yer}},
  \bibinfo{journal}{Phys. Rev. A} \textbf{\bibinfo{volume}{62}},
  \bibinfo{pages}{052304} (\bibinfo{year}{2000}).

\bibitem[{\citenamefont{Long}(2001)}]{long01a}
\bibinfo{author}{\bibfnamefont{G.~L.} \bibnamefont{Long}},
  \bibinfo{journal}{Phys. Rev. A} \textbf{\bibinfo{volume}{64}},
  \bibinfo{pages}{022307} (\bibinfo{year}{2001}).

\bibitem[{\citenamefont{Grover}(2005)}]{grover05}
\bibinfo{author}{\bibfnamefont{L.~K.} \bibnamefont{Grover}},
  \bibinfo{journal}{Phys. Rev. Lett.} \textbf{\bibinfo{volume}{95}},
  \bibinfo{pages}{150501} (\bibinfo{year}{2005}).

\bibitem[{\citenamefont{Pipes}(1958)}]{pipes58}
\bibinfo{author}{\bibfnamefont{L.~A.} \bibnamefont{Pipes}},
  \emph{\bibinfo{title}{Applied mathematics for engineers and physicists}}
  (\bibinfo{publisher}{Mc-Graw-Hill Book Company, Inc.},
  \bibinfo{address}{Toronto}, \bibinfo{year}{1958}), \bibinfo{edition}{2nd} ed.

\bibitem[{\citenamefont{Sprung et~al.}(1993)\citenamefont{Sprung, Wu, and
  Martorell}}]{sprung93}
\bibinfo{author}{\bibfnamefont{D.~W.~L.} \bibnamefont{Sprung}},
  \bibinfo{author}{\bibfnamefont{H.}~\bibnamefont{Wu}}, \bibnamefont{and}
  \bibinfo{author}{\bibfnamefont{J.}~\bibnamefont{Martorell}},
  \bibinfo{journal}{Am. J. Phys.} \textbf{\bibinfo{volume}{61}},
  \bibinfo{pages}{1118} (\bibinfo{year}{1993}).

\end{thebibliography}
\end{document}